\documentclass[twocolumn,prl,a4paper,floatfix,preprintnumbers,amsmath,amssymb]{revtex4}

\usepackage{amsmath}
\usepackage{graphicx}

\usepackage{graphicx}
\usepackage{dcolumn}
\usepackage{bm}

\newcommand{\ud}{\mathrm{d}}

\newcommand{\be}{\begin{equation}}
\newcommand{\ee}{\end{equation}}

\begin{document}

\title{Zeno Dynamics and Distinguishability of Quantum States}

\author{Augusto Smerzi}
\affiliation{ BEC-CNR-INFM and Dipartimento di Fisica, Universit\`a di Trento, I-38050 Povo, Italy}

\date{\today}
                      
\begin{abstract} 
According to the quantum Zeno effect, the frequent observations of a system can dramatically slow down its dynamical evolution.
We show that the Zeno dynamics is the result of projective measurements among quantum states which are indistinguishable. 
The physical time scale of the problem is provided by the Cramer-Rao lower bound, which measure
the distinguishability of states along a path in the Hilbert space. We finally show that the Zeno dynamics with particle entangled states might 
require quite smaller measurement intervals than classically correlated states, 
and propose a realistic interferometric experiment to test the prediction. 
\end{abstract}
Ê
\maketitle

{\it Introduction}.
Watching a quantum system can freeze its dynamical evolution \cite{misra77}. This vindicates 
one of the classical Zeno paradoxes arguing that a flying arrow cannot really move since at each instant of time
it occupies the same region of space \cite{zeno}. 
In the quantum world, the irreducible perturbations created by the observations can indeed 
repeatedly bring back the arrow to its initial position.
The main result of this manuscript is to show that this is possible when the different 
quantum states of the arrow are statistically indistinguishable. 

The quantum Zeno effect (QZE) has raised and it continue to gather widespread interest \cite{facchi08} mainly because of two reasons.
For its foundational implications about the nature of a ``quantum measurement" \cite{home97},
and for its technological applications in quantum information since it can be exploited to 
preserve decoherence free regions where to build up quantum computation protocols 
\cite{viola98,franson04,hosten06}.
Various aspects of QZE have been experimentally demonstrated with ions \cite{itano90}, 
polarized photons \cite{kwiat95}, cold atoms \cite{fischer01},
dilute Bose-Einstein condensed gases \cite{streed06}.
The paradoxical (or, at least, surprising) nature of the problem,
however, had sometimes obscured its physical significance \cite{home92}.
The fact is that Quantum Mechanics allows for the existence of a quantum Zeno time scale $\tau_{qz}$. If the interval between consecutive
measurements is smaller than $\tau_{qz}$ the dynamics is
significantly slowed down or even asymptotically halted. 

{\it Zeno dynamics and distinguishability of quantum states}.
In this manuscript we show that the quantum Zeno effect is the physical
consequence of a
statistical indistinguishability of quantum states.
The concept of statistical distance (or distinguishability) in the Hilbert space  
was introduced by Wootters as the number of states
that can be {\it physically discriminated by a measuring device }
with a sufficiently large number $m$ of measurements \cite{wootters81}. 
If the quantum states differ by the value of a parameter as, for instance, a phase shift due to the interaction with some external perturbation or an elapsed time, 
the smallest path interval of two neighboring distinguishable states is 
\be \label{cr}
\tau_{sd} = \frac{2}{\sqrt{m} \sqrt{F} }.
\ee
The leading role in the theory is played by the Fisher information $F$.
Its explicit mathematical expression, see Eq.(\ref{f}) below, depends on the state, the generator of the quantum path and the choice of the measured observable. 

As will demonstrate, the Fisher information plays a central role in the quantum Zeno dynamics.
The survival probability $P(t)$ 
that the state at time $t$ remains identical to its initial value 
after a large number of projective measurements $m$ can be written as: 
\be \label{sp}
P(t) \simeq 1 - \frac{F}{4 m}~t^2  = 1 - \left({\tau \over \tau_{qz}} \right)^2 
\ee
where $\tau = t/ m$ is the time interval between {\it two} consecutive observations.
The Eq.(\ref{sp}) quite generally describes the quantum Zeno dynamics from the case of 
simple projective measurements to more sophisticated ``bang-bang" and continuous measurements controls. 
The ratio $\tau / \tau_{qz}$ is the small parameter of the theory, 
with the quantum Zeno time scale being
\be \label{qze}
\tau_{qz} =  \tau_{sd} 
\ee
The Eq.s(\ref{sp},\ref{qze}) summarize the main results of this manuscript.

{\it The quantum Zeno effect}.
We now recall the general argument leading to the quantum Zeno effect (QZE). 
For sake of clarity, we first consider the simple dynamical 
evolution of a pure state. The generalization
to arbitrary density matrices and subspace projections will be discussed below.

Consider an Hamiltonian $ \mathcal{\hat H}$ driving an initial state $|\psi_0\rangle$
to $|\psi(t)\rangle = e^{-i \mathcal{\hat H} t} |\psi_0\rangle$ after the time $t$. 
The survival probability to find the evolved state in its initial configuration is $P(t)= |\langle \psi_0 |\psi(t)\rangle|^{2}$.
Suppose now that during the dynamics the system is observed $m$ times at equal intervals $\tau = t / m$
with projective measurements $\hat \Pi = |\psi_0\rangle \langle \psi_0|$. 
The survival probability becomes
$P(t)=  |\langle \psi_0| e^{-i \mathcal{\hat H} \tau} |\psi_0\rangle \langle \psi_0|...  |\psi_0\rangle \langle \psi_0| e^{-i \mathcal{\hat H} \tau} |\psi_0\rangle|^2=  |\langle \psi_0| \psi(\tau)\rangle|^{2 m}$. In the limit of a small interval $\tau$
among consecutive measurements we can 
expand $e^{-i \mathcal{H} \tau}$ up to second order and have:
\be \label{sp1}
P(t) =  |\langle \psi_0|\psi(\tau)\rangle|^{2 m} = 1 - m~ \Delta^2 \mathcal{\hat H}~ \tau^2 + O(\tau^4)
\ee
The initial state does not significantly evolve with time,  $P(t) \simeq 1$,
if $m~ \Delta \mathcal{\hat H}~ \tau^2 \ll1 $.
The Zeno dynamics is the consequence of the quadratic short time evolution of the survival probability \cite{facchi08}.

The extension of this result to general states and projections on Zeno subspaces is less straightforward and has been introduced 
in \cite{facchi02}. 
The unitary dynamical evolution of the system is given by $\rho(t) = e^{-i \mathcal{\hat H} t} \rho_0 e^{i \mathcal{\hat H} t}$. 
The measurement consists on a projection $\hat \Pi$
which does not commute with the hamiltonian, $[\hat \Pi, \mathcal{\hat H}]\neq 0$ and is defined in the eigenspace 
given by $\mathcal{\hat H}_{\Pi}=\hat \Pi~\mathcal{\hat H}~\hat \Pi$. 
The initial density matrix $\rho_0$ is defined in $\mathcal{\hat H}_{\Pi}$,
so that $\rho_0 = \hat \Pi \rho_0 \hat \Pi$, and 
$Tr[\rho_0 \hat \Pi] = 1$.

As in the previous case of pure states, we consider
$m$ observations at equal time intervals $\tau = t / m$.
The survival probability to find the system in the subspace $\mathcal{H}_\Pi$ is given by:
\be \label{sp2}
P(t) = Tr[V_m(t) \rho_0 V_m^\dagger(t)]   
\ee
where $V_m(t) = \big(\hat \Pi e^{-i \mathcal{\tilde H} t/m} \hat \Pi \big)^m$. In the limit $m \to \infty$, $P(t) \to 1$  \cite{facchi02}.
For our purposes, we need the expand Eq.(\ref{sp2}) for small $\tau$:
\begin{equation}
\begin{aligned}
&P(t) \simeq Tr[(1 + i \mathcal{\hat H} \tau - {1 \over 2} \mathcal{\hat H} \tau^2)^m
\hat \Pi (1 - i \mathcal{\hat H} \tau - {1 \over 2} \mathcal{\hat H} \tau^2)^m ~\rho_0]& \\
&\simeq Tr[\hat \Pi + m~(\hat \Pi \mathcal{\hat H}^2 \hat \Pi + \hat \Pi \mathcal{\hat H} \hat \Pi \mathcal{\hat H} \hat \Pi)~\tau^2 ~\rho_0]&  \\
&= 1 - m~  \Delta^2 \hat{\rm H}~ \tau^2&
\end{aligned}
\label{sp3}
\end{equation} 
where
\be \label{eh}
\hat {\rm H} = \mathcal{\hat H}-\hat \Pi \mathcal{\hat H} \hat \Pi
\ee
The initial state remains in the Zeno subspace $\mathcal{\hat {H}}_\Pi$
when $m~ \Delta {\hat {\rm H}}~ \tau^2 \ll1 $.
Needless to say, the pure state case previously discussed is recovered with
$\rho_0 = |\psi_0\rangle \langle \psi_0|$ and $\hat \Pi = |\psi_0\rangle \langle \psi_0|$.

{\it Quantum Zeno and Fisher information}.
We now prove the relation between quantum Zeno, Fisher information and statistical distance.
The number of states that can be statistically distinguished with $m$ measurements along the path parametrized by $t$ and
connecting the two states $\rho_0$ and $\rho(t) = e^{-i \mathcal{\hat H} t} \rho_0 e^{i \mathcal{\hat H} t}$ is given by
\be \label{sd}
N_{ds} = {\sqrt{m} \over 2} \int_0^t~\sqrt{F(\tau)}~d \tau  
\ee
when $m \gg 1$ \cite{wootters81}.
The Fisher information is given by
\be \label{f}
F(\tau) = \int \ud \eta \frac{1}{\mathcal{P}(\eta|\tau)} \left(\frac{\ud \mathcal{P} (\eta|\tau)}{\ud \tau}\right)^2
\ee
where $\mathcal{P} (\eta|\tau) = \mathrm{Tr}[\hat{M}(\eta) \hat{\rho}(\tau)]$ is the likelihood, i.e., the conditional probability to 
obtain from the measurement a value $\eta$ for a given $\tau$.
Eq.(\ref{f}) depends on $\rho(t)$, the generator of the transformation $\mathcal{\hat H}$,
and the choice of the observable, which is quite generally provided by a complete set of Hermitian operators $\hat M(\eta)$, 
where $\eta$ labels the result of the measurement and $\int \ud \eta~ \hat{M}(\eta) =\hat{I}$ (identity). 

How is the Fisher information related with quantum Zeno?
With projective measurements there are only two possible outputs,
which we call ``yes" and ``no", occurring with probability 
$\mathcal{P}(\rm{yes}|\tau) = Tr[\hat \Pi~e^{-i \mathcal{\hat H} \tau} \rho_0~e^{i \mathcal{\hat H} \tau}]$ and 
$\mathcal{P}(\rm{no}|\tau)= 1- \mathcal{P}(\rm{yes}|\tau)$, respectively.
Eq.(\ref{f}) gives
\begin{equation}
\begin{aligned}
F(\tau) = \left(\frac{ \mathcal{P}(\rm{yes}|\tau)}{\partial \tau}\right)^2 ~{\frac{1}{ \mathcal{P}(\rm{yes}|\tau) [1 - \mathcal{P} (\rm{yes}|\tau)]}}. 
\end{aligned}
\label{f1}
\end{equation} 
Since we are considering small time intervals $\tau$ between measurements, 
we have  $\mathcal{P}(\rm{yes}|\tau)  = 1 - \Delta^2 \hat{\rm H}~ \tau^2 + O(\tau^4)$ (cfr. Eq.(\ref{sp3})).
We finally obtain
\begin{equation}
\begin{aligned}
F =
 4~\Delta^2 {\hat {\rm H}} + O(\tau^4)
\end{aligned}
\label{f2}
\end{equation} 
with ${\hat {\rm H} }$ given by the Eq.(\ref{eh}). Notice that the Fisher is independent from the value of $\tau$ 
up to $O(\tau^4)$. By replacing Eq.(\ref{f2}) in Eq.(\ref{sp3}) we finally find the general expression for the survival probability
Eq.(\ref{sp}). The Eq.(\ref{qze}) follows from the Eq.(\ref{sd}).

{\it Discussion}.
There are two facets of Eq.s(\ref{sp},\ref{qze}) which deserve some discussion. 
The first one is that we can set strong bounds on $\tau_{qz}$ for states which are entangled or classically correlated.
The Fisher information indeed plays an important role in the theory of entanglement.
Consider a state of of $N$ q-bits. It exists a class of entangled states, recognized by the condition $N < F \leq N^2$,
which can provide sub-shot noise sensitivity in interferometric phase estimation problems \cite{pezze09}. 
On the other hand, with classically correlated states $ F \leq  N$. Therefore, because of Eq.(\ref{sp}), the Zeno dynamics with
entangled states might require quite smaller intervals between measurements with respect to 
separable states. This can be important for the creation of decoherence free regions in quantum computation
applications. We will propose a realistic experimental test of this effect.

The second comment is that Eq.s(\ref{sp},\ref{qze}) allow to interpret the quantum Zeno as a parameter estimation 
problem. Generally speaking, indeed, the projective measurements can be considered as quantum metrology
attempts to estimate $\tau$. This is the length separating two states among consecutive observations.
To discriminate $\tau$ from the noise is equivalent to recognize the two states as different. 
The highest sensitivity of the estimation (smallest mean-square fluctuation) allowed by quantum mechanics
is provided by the Cramer-Rao lower bound
$\Delta \tau_{cr} = {1 \over  \sqrt{m F}}$ 
and the quantum Zeno time can be therefore written as 
\be \label{ztcr}
\tau_{qz}  = 2 \Delta \tau_{cr}.
\ee 
Let's follow the Zeno dynamics at fixed intervals $\tau$.
The first measurement ($m =1$) is done at the time $t = \tau$. If the ``signal" $\tau$ is smaller than the ``noise" $\Delta \tau_{cr} = 1 / \sqrt{F}$,
the evolved state is indistinguishable from its initial value. In this case the perturbation due to the
projective measurement {\it brings back} the evolved state to its starting configuration. 
The successive measurements steadily increase the sensitivity, namely, our confidence that the two states are actually distinct.   
However, if after $m$ measurements, at the time $t = m \tau$, the path interval $\tau$ is still smaller than $\Delta \tau_{cr} \sim 1 / \sqrt{m F}$,
the survival probability remains $P(t) \simeq 1$ and the dynamics is Zeno.
By still increasing the number of measurements, inevitably at some point $\tau \sim \Delta \tau_{cr}$ and the two states are finally
recognized as different. At this moment the dynamics ceases to be Zeno.
Mathematically, this is because the expansion bringing to Eq.(\ref{sp}) breaks down.
At larger times, when $\tau > \Delta \tau_{cr}$, the anti-Zeno effect \cite{antizeno} might accelerate the dynamics. 
Asymptotically in $m$,  
$\Delta \tau_{cr} \to 0$ and the survival probability eventually becomes independent from our choice to perform the 
projective measurements or not.

To summarize:
i) the Cramer-Rao lower bound Eq.({\ref{cr}) provides the natural time scale of the problem; 
ii) the dynamics is slowed down when the projective measurements are performed in intervals so small
that the two neighboring states are not distinguishable;
iii) the minimal quantum Zeno interval can be quite different for states which are separable or entangled. 
We now elaborate on this last point by considering a simple example.

{\it Quantum Zeno and entanglement}.
Consider the dynamics of $N$ q-bits governed by
the Hamiltonian $H = \hbar \omega \sum_{l=1}^{N} \hat{\sigma}_z^{(l)}$, where $\sigma_z^{(l)}$ is the $z-$ component of the Pauli
operators.
If the state is classically correlated, namely, it can be written as a convex combination of separable states: 
\be \label{rho}
\hat{\rho}_{\mathrm{sep}} = \sum_k p_k \, \hat{\rho}^{(1)}_k \otimes \hat{\rho}^{(2)}_k \otimes ... \otimes \hat{\rho}^{(N)}_k,
\ee
then the Fisher information is bounded by $F \leq N (\hbar \omega)^2$. Therefore, for separable states 
\be \label{sep}
\tau_{qz} \geq \frac{2}{\hbar \omega~\sqrt{m~N}}
\ee
On the other hand, the highest possible value of the Fisher is $F \leq N^2 (\hbar \omega)^2$. The bound can be saturated with
maximally entangled states $|\psi_0\rangle = \left(|\uparrow^N\rangle + |\downarrow^N\rangle \right) / \sqrt{2}$, where the 
$ |\uparrow\rangle$,  $|\downarrow\rangle$ are the eigenstates of $\hat \sigma_z$.

Therefore, a sufficient condition for the presence of 
entanglement in a quantum state is that its corresponding Fisher information $F > N (\hbar \omega)^2$ \cite{pezze09}. This class of states 
requires a time interval to create the Zeno dynamics
\be \label{ent}
\frac{2}{\hbar \omega~N \sqrt{m}} \leq \tau_{qz} < \frac{2}{\hbar \omega~\sqrt{N m}} .
\ee
which can be smaller up to a factor  $\sqrt{N}$
than the Zeno time of separable states.

The predictions Eq.s(\ref{sep},\ref{ent}) can be tested experimentally with Mach-Zehnder (MZ) interferometers.
A MZ is mathematically described by a unitary evolution which rotates the initial state by an angle corresponding to the phase
shift $\theta$ applied between the two arms of the interferometer:
\be \label{mz}
|\psi(\theta) \rangle = e^{- i \hat J_y~\theta}  |\psi_{inp} \rangle 
\ee
The generator of the phase shift $\theta$ is the $y$-component of the
pseudo angular momentum $\hat{\vec{J}} = (\hat{J}_x, \hat{J}_y, \hat{J}_z)$, with
$\hat J_x = ({\hat a}^\dagger \hat b + {\hat b}^\dagger \hat a) / 2,
\hat J_y = ({\hat a}^\dagger \hat b - {\hat b}^\dagger \hat a) / 2 i,
\hat J_z = ({\hat a}^\dagger \hat a - {\hat b}^\dagger \hat b) / 2$
and $\hat a, \hat b$ destruction operators of the two input modes \cite{yurke86}. 
A typical observable is the relative number of particles measured at the two output ports, $(\hat{N_c} - \hat{N_d})/2 = \hat{J}_z$. 
Their quantum mechanical average calculated over an ensemble of experiments repeated under identical conditions, is: 
\begin{equation}
\begin{aligned}
\langle \hat{J}_z \rangle_{out} = \langle \hat{J}_z \rangle_{inp}~ \cos{\phi} - \langle \hat{J}_x \rangle_{inp}~ \sin{\phi} 
\end{aligned}
\label{mz1}
\end{equation} 
Let's choose as input of the MZ a product state $|\psi_{inp}\rangle  = |\psi_a \rangle | \psi_b \rangle$. 
We consider $m$ Mach-Zehnder interferometers sequentially connected
so that the particles leaving the output ports $\{c_j, d_j\}$ of the $j^{th}$ interferometer,
enter the input ports $\{a_{j+1}, b_{j+1}\}$ of the next one as in Fig.(\ref{seq_mz},a).
In each interferometer we apply a phase shift $\theta / m$. 
The rotation of the initial state is now simply given by 
\be \label{mz}
|\psi(\theta) \rangle = \prod_{i=1}^m e^{- i \hat J_y~  \theta / m}  |\psi_{inp} \rangle 
\ee 
Therefore, rather trivially, the global phase shift experienced by the particles crossing all the interferometers is $\theta$ and
the average relative number of particles eventually detected at the outputs of the last interferometer is still given by Eq.(\ref{mz1}). 

To study the Zeno dynamics, we
cut the connections at the output ports $c_1, c_2, ...c_m$. We can leave undetected the particles which might exit those ports. 
The input ports $a_1, a_2...a_m$ are instead injected with the initial state $|\psi_a \rangle$, see Fig.(\ref{seq_mz},b).
How many particles are detected in average at the output of the last interferometer? if 
the number $m$ of interferometers is sufficiently large (the phase shift $\theta /m$ suffficiently small), than the average 
number of particles detected in outout is equal to the average number of particles entering the first MZ,
$\langle \hat{J}_z \rangle_{out} \simeq  \langle \hat{J}_z \rangle_{inp}$ and the evolution is Zeno.

How large has to be $m$ ?
Let's consider first the case of classically correlated input states.
With a coherent state having an average number of particles $\bar{N} = |\alpha|^2$ 
injected in one of the two input ports while leaving the vacuum in the other one,
$|\psi_{inp}\rangle  = |0\rangle_a | \alpha \rangle_b$, Eq.(\ref{sep}) 
gives $m > \bar{N}$.
Entanglement can change the scenario.
We now inject the previously unused ports with a Foch state
of $N$ particles, so that $|\psi_{inp}\rangle  = |N \rangle_a |\alpha \rangle_b$.
This is a particle entangled state \cite{hyllus} which provides an interferometric sub shot-noise phase sensitivity when $N > 0$,
up to the Heisenberg limit at the optimal choice $N \sim \bar{N}$  \cite{pezze}.
In the latter case,
the number of MZ needed to observe the Zeno dynamics is $m > \bar{N}^2$.
Therefore, the phase shift $\theta / m$ in each interferometer has to be smaller by a factor $1 / \sqrt{\bar{N}}$ for usefully
entangled states with respect to classically correlated states. 
These are experiments which can be realistic performed within the current state of the art. 
\begin{figure}[!t]
\begin{center}
\includegraphics[scale=0.5]{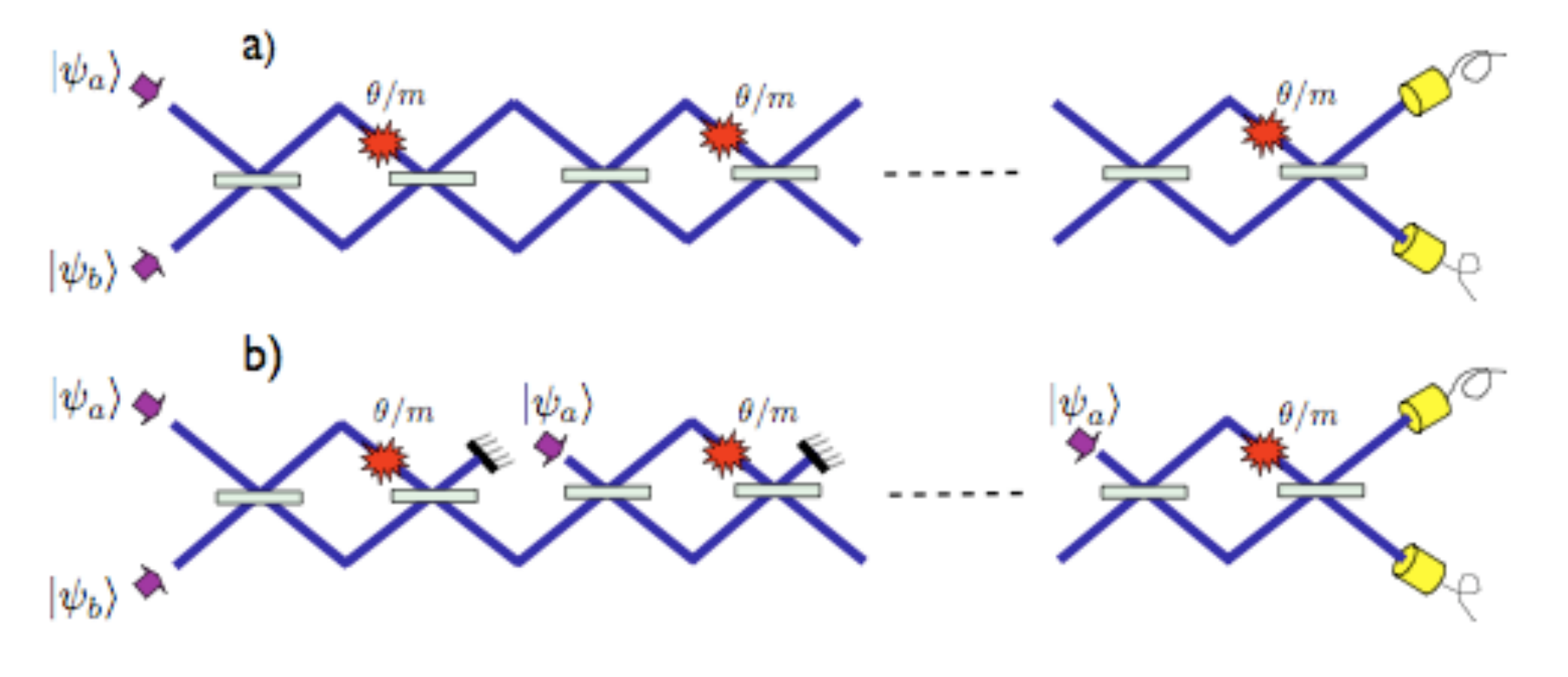}
\end{center}
\caption{\small{
Two different setups consisting of a sequence of $m$ Mach-Zehnder interferometers. 
The phase shift in each MZ is $\theta / m$. The input state is 
the product $|\psi_a \rangle |\psi_b \rangle$.
The detectors count the number of particles in output. 
a) The MZ are sequentially connected so that the output of each interferometer becomes the input of the next one.
b) In this configuration only one output port becomes the input of the next interferometer. The remaining output port
remains disconnected, while the corresponding input port of the next MZ is 
injected with $| \psi_a \rangle$. With a sufficiently large number $m$ of interferometers, the evolution is quantum Zeno.} }
\label{seq_mz} 
\end{figure}

{\it Some final further remarks for the case of pure states}.
Quite generally, the Fisher information depends on the choice of the observable. It is possible to calculate, however, the highest vale of the Fisher
obtained with an optimal choice of the measurement apparatus, the one that can better discriminate neighboring states. 
This is referred  in the literature as quantum Fisher \cite{braunstein94}.
For the case of pure states, the quantum Fisher is given by $ |\langle \psi(t)|\psi(t+\delta t)\rangle|^2 = 1 - F_q~(\delta t)^2$, and is
precisely $F_q = 4~ \Delta^2 \mathcal{\hat H}$. Therefore, the survival probability can be written as:
\be \label{sp4}
P(t) \sim e^{-( \tau / 2 \Delta t_{qcr})^2}
\ee
where $\Delta t_{qcr} = 1 / \sqrt{m~F_q}$ is the quantum Cramer-Rao bound. It is possile to define the quantum Fisher 
also for impure density operators \cite{braunstein94}. However, in this case the Fisher entering in Eq.(\ref{sp}) is not the quantum Fisher. 
Notice also that the Cramer-Rao can be written as an uncertainty relation. For pure states we simply have
$\Delta t_{qcr}~ \Delta \mathcal{\hat H} \ge 1 / 2 \sqrt{m}$. This relation has of course a different meaning than
an Heisenberg uncertainty relation since it refers to fluctuations of a measured parameters rather than Hermitian operators \cite{milburn}. 
In particolar, the parameter can be estimated with arbitrary precision, by just increasing the number of measurements $m$.
The fact that in the quantum Zeno the physics is related to a parameter based uncertainty relation instead of
an Heisenberg uncertainty relation is a different way to summarize our results.

{\it The quantum Zeno paradox and conclusions}.
The frequencies of measurement outcomes 
can in general deviate from the exact quantum mechanical probabilities. This simply because of
statistical fluctuations of a finite sample of data. Therefore, two slightly different systems can actually be 
indistinguishable with respect to a finite number of measurements 
if the difference of the respective quantum mechanical probabilities are smaller than the fluctuations of the 
frequencies \cite{wootters81}.
In the classical world, indistinguishability is just the consequence of ignorance, without further complications.
Not surprisingly, this is not the case in the quantum word.
If the two neighboring states are too close to be distinguished with a reasonable confidence, then the projective measurement brings back the evolved state to the projected one. This is the bottom line of the quantum Zeno paradox: an eternal returning of the arrow to the bow.
To conclude, we notice that there are different technologies which are based on efficiently distinguish quantum states. Our results can therefore
be extended and applied in various contexts as, for instance, in quantum control theories, when searching the optimal path to generate a target quantum state
\cite{calarco}, in the conditions for adiabatic approximations \cite{boixo} and applications in adiabatic quantum computation \cite{farhi}, 
in the estimation of the speed limits of quantum computation \cite{levitin}.

{\it Acknowledgements}.
We gratefully acknowledge discussions with T. Calarco and P. Hyllus.

\end{document}